\documentclass[prl,notitlepage,twocolumn]{revtex4}

\usepackage{amsmath}
\usepackage{amssymb}
\usepackage{bm}
\usepackage{graphicx}
\usepackage{times}
\usepackage{xcolor}
\usepackage{float}
\usepackage{gensymb}

\usepackage{pdfcomment}
\pdfcommentsetup{draft} 
\pdfcommentsetup{color={1.0 1.0 0.0}, open=true}

\usepackage{color}
\usepackage{ulem}

\begin{document}

\title{Meta-optics and bound states in the continuum}

\author{Kirill Koshelev$^{1,2}$}
\author{Andrey Bogdanov$^{2}$}
\author{Yuri Kivshar$^{1,2}$}
\affiliation{$^1$Nonlinear Physics Centre, Australian National University, Canberra ACT 2601, Australia}
\affiliation{$^2$ITMO University, St. Petersburg 197101, Russia}

\begin{abstract}
We discuss the recent advances in meta-optics and nanophotonics associated with the physics of bound states in the continuum (BICs).
Such resonant states appear due to a strong coupling between leaky modes in optical guiding structures being supported by subwavelength high-index dielectric Mie-resonant nanoantennas or all-dielectric metasurfaces.  First, we review briefly very recent developments in the BIC physics in application to isolated subwavelength particles. We pay a special attention to novel opportunities for {\it nonlinear nanophotonics} due to the large field enhancement inside the particle volume creating the resonant states with high-quality (high-$Q$) factors, the so-called quasi-BIC, that can be supported by the subwavelength particles.  Second, we discuss novel applications of the BIC physics to all-dielectric optical metasurfaces with broken-symmetry meta-atoms when tuning to the BIC conditions allows to enhance substantially the $Q$ factor of the flat-optics dielectric structures.  We also present the original results on {\it nonlinear high-$Q$ metasurfaces} and predict that the frequency conversion efficiency can be boosted dramatically by smart engineering of the asymmetry parameter of dielectric metasurfaces
 in the vicinity of the quasi-BIC regime.
\end{abstract}

\maketitle

\section{Introduction}

The study of resonant dielectric nanostructures with large refractive index is a new research direction in nanoscale optics and metamaterial-inspired nanophotonics~\cite{kuznetsov2016optically,staude,kruk2017functional}. Many concepts of all-dielectric resonant nanophotonics are driven by the idea to employ subwavelength dielectric nanoparticles with Mie resonances as ``meta-atoms'' for creating highly efficient optical metasurfaces and metadevices~\cite{nmat}. Precise engineering of the optical resonances results in many fascinating phenomena such as perfect reflection, broadband transmission, complete control of phase and polarization, imaging with subwavelength resolution~\cite{Moitra2015,Faraon2015,Shalaev2015,Yu2015}. The pronounced resonant properties of high-index dielectric nanoparticles along with low-cost fabrication and compatibility with planar technology make high-index nanoscale structures prospective candidates to complement or even replace different plasmonic components in a range of potential applications. Moreover, many concepts which had been developed for plasmonic structures but could not be fully employed due to strong losses of metals, can now be realized based on low-loss dielectric structures.

High-index dielectric meta-atoms can support both electric and magnetic Mie resonances in the visible and mid-IR spectral ranges, which can be tailored by the nanoparticle geometry. To emphasize the importance of optically-induced magnetic response, the field of all-dielectric resonant nanophotonics is often termed as {\it meta-optics}. The study of Mie-resonant silicon nanoparticles has recently received considerable attention for applications in nanophotonics and metamaterials~\cite{yuri} including optical nanoantennas, wavefront-shaping metasurfaces, and nonlinear frequency generation. Importantly, the simultaneous excitation of strong electric and magnetic Mie-type dipole and multipole resonances can result in constructive or destructive interferences with structured beam~\cite{oe_wei}, and it may also lead to the resonant enhancement of magnetic fields in dielectric nanoparticles that could bring many novel effects in both linear and nonlinear regimes.

The physics and applications of all-dielectric resonant nanophotonics could be extended substantially by employing the concept of {\it bound states in the continuum} (BICs), which represent localized states with energies embedded in the continuous spectrum of radiating waves. BICs were predicted as a mathematical curiosity in quantum mechanics long time ago~\cite{von1929uber}. In spite of the fact that the system proposed in that work has never been implemented experimentally, the beautiful physics of BIC was widely used, first in atomic physics~\cite{Fonda1963,Stillinger1960,fri_win} and then in acoustics and hydrodynamics~\cite{Ursell1951,Cumpsty1971,Parker1989}. For last years, BICs have been attracting very broad attention in photonics~\cite{BIC_review}, primarily because they provide a simple way to achieve very large quality factors ($Q$ factors) for photonic crystals, metasurfaces, and even subwavelength isolated resonators.

The key idea underlying bound states in the continuum is vanishing coupling between the resonant mode and all radiation channels of the surrounding space. The coupling coefficient could vanish due to the symmetry reason when the spatial symmetry of the mode is incompatible with the symmetry of the of outgoing radiating waves. Such kind of BIC is called {\it symmetry-protected}. It appears in a variety of photonic structures such as gratings and metasurfaces, waveguide arrays, waveguides coupled to resonators, and many other~\cite{Paddon2000,Tikhodeev2002,Shipman2005,Bulgakov2014,Bulgakov2017,Belyakov2018}. In contrast to the symmetry-protected BIC, there is a so-called {\it accidental BIC} forming due to accidental vanishing of the coupling coefficients to the radiation waves via continuous tuning of one or several system parameters. The simplest example of accidental BIC is a Fabry-Perot-type BIC. Such a state is formed between two mirrors resonantly reflecting the waves placed at a proper distance providing the phase shift multiple of $2\pi$ after the round-trip~\cite{Marinica2008,Ndangali2010,Bulgakov2015}. Some of accidental BICs could be explained in terms of destructive interference of two (or more) leaky waves, whose radiation is tuned to cancel each other completely. This mechanism is known as {\it Friedrich-Wintgen scenario}~\cite{fri_win}. There are more specific examples of BIC arising, for example, in anisotropic~\cite{GomisBresco2017}, Floquet~\cite{Zhong2017} or PT-symmetric systems~\cite{Longhi2014, Kartashov2018}. Today the structures with optical BICs are successfully used for filtering, lasing, sensing and Raman spectroscopy~\cite{Foley2014,Liu2017,Romano2018,BIC_nature,kuznetsov_lasing,sensing,SERS}. Application of BICs for nonlinear optics, twisted light and light-mater interaction is under active study~\cite{Belyakov2018, Bulgakov2014,Bulgakov2015,Krasikov2018,Bulgakov2017_1,Bulgakov2017_2,BIC_exciton,Poddubny2018}.
\begin{figure*}[t]
   \centering
\centering\includegraphics[width=0.9\linewidth]{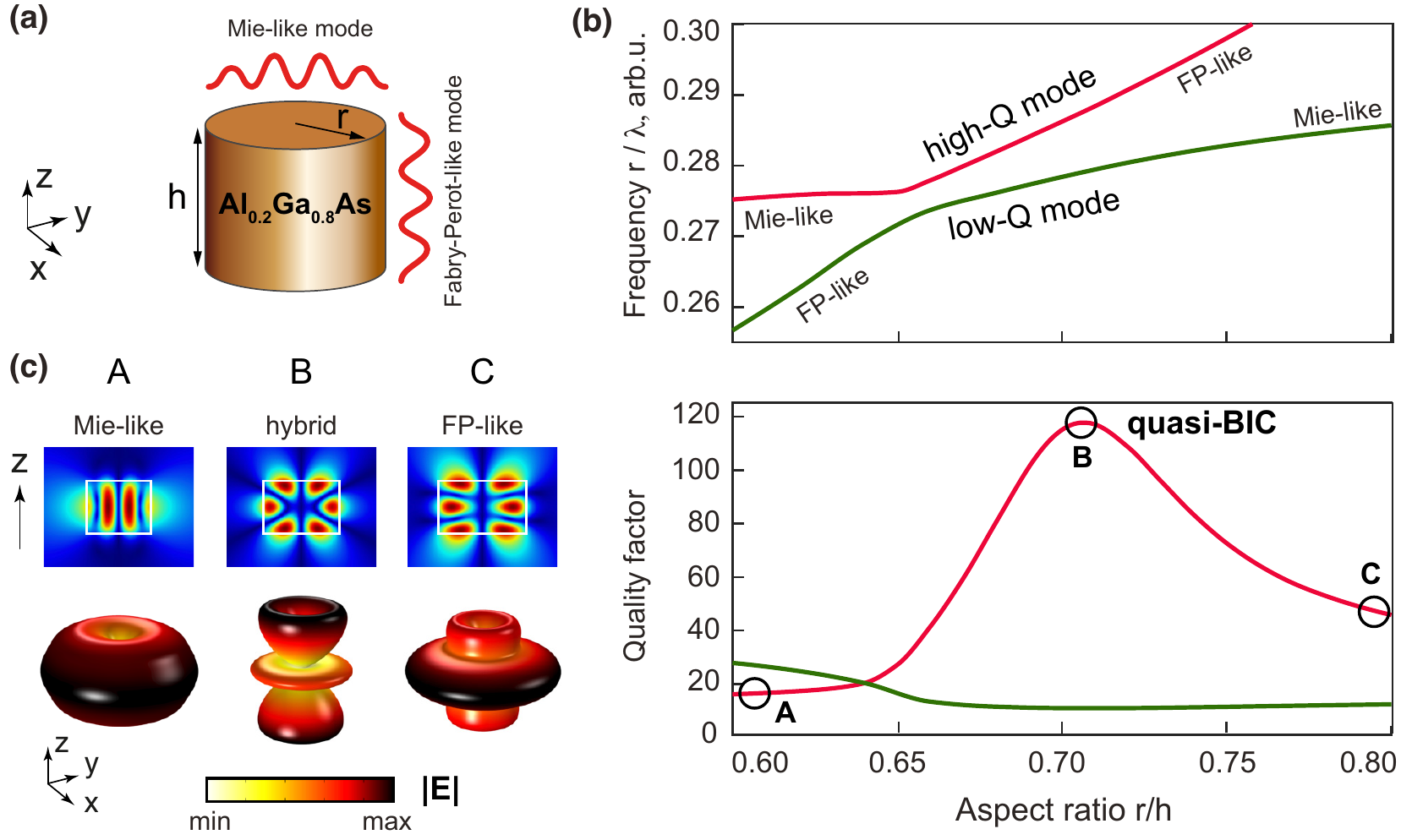}
 \caption{  {\bf Quasi-BICs in isolated subwavelength nanoparticles.}
 (a) Geometry of a AlGaAs cylindrical nanoparticle resonator with radius $r$ and height $h$. (b) Strong coupling of two resonator modes resulting in the emergence of a quasi-BIC~\cite{prl_kirill}. Upper panel: Mode frequencies vs. aspect ratio of the nanoparticle. Lower panel: Mode $Q$ factor vs. aspect ratio. Points A, B, and C mark there regimes of the mode coupling, before and after the quasi-BIC supercavity condition is satisfied. (c) Upper panel: Near-field distributions of the electric field at the points A, B, and C.  Lower panel: Transformation of the radiation pattern for the high-$Q$ mode while passing the avoided resonance crossing~\cite{bogdanov,bic_wei}. The quasi-BIC supercavity mode condition corresponds to the magnetic octupole mode.
}
\label{fig1}
\end{figure*}

In practice, infinitely high Q factors of BICs are limited by a finite size of samples, material absorption, structural disorder and surface scattering~\cite{Sadrieva2017, Belyakov2018}. As a result, BICs are transformed into states with a giant $Q$ factor often treated as {\it quasi-BICs}. Remarkably, quasi-BICs can form even in a single subwavelength high-index dielectric resonator by tuning the structure parameters into the so-called {\it supercavity} regime~\cite{prl_kirill,rybin,Taghizadeh2017}. In connection with broader applications, many metasurfaces composed of arrays of dissimilar meta-atoms with a broken in-plane symmetry can support high-$Q$ resonances directly associated with the concept of quasi-BICs~\cite{new_PRL}. This includes, in particular, broken-symmetry Fano dielectric metasurfaces for enhancement of nonlinear effects~\cite{brener} and recently reported biosensing with pixelated dielectric metasurfaces~\cite{science}. Currently, quasi-BICs in all-dielectric structures have been shown to be advantageous for a wide span of applications, including lasing~\cite{BIC_nature, kuznetsov_lasing}, active photonics~\cite{active}, polaritonics~\cite{BIC_exciton} and biophotonics~\cite{sensing,SERS}.

The aim of this paper is twofold.  First, we summarize briefly very recent advances in the physics of bound states in the continuum in application to isolated subwavelength particles. We pay a special attention to novel opportunities for nonlinear nanophotonics due to the fact that high-$Q$ states (quasi-BICs) supported by subwavelength particles are associated with large energy concentration inside the particle volume. Second, we discuss the application of these ideas to the physics of optical metasurfaces with broken-symmetry meta-atoms when tuning to the BIC conditions allows to enhance substantially the $Q$ factor of the structure.  In this section, we also present the original results about nonlinear high-$Q$ metasurfaces and predict that the frequency conversion efficiency can be boosted dramatically by smart engineering the asymmetry parameter of dielectric metasurfaces in the vicinity of the quasi-BIC.

\section{Isolated nanoparticles}

To date, the nanoscale confinement of light and high $Q$ factor of a resonator are crucial for many applications in photonics. Conventional ways to enlarge the $Q$ factor are to increase the size of a resonator, for example, by exploiting whispering gallery modes and cavities in photonic crystals, or to arrange several resonators in space and excite collective modes. Although such techniques allow for light trapping with reduced energy losses, the size of the resonating system is still large compared to the operating wavelength. Recently, a completely different approach was proposed based on the concepts of BICs and supercavity modes which allows for subwavelength light confinement in all-dielectric high-$Q$ resonators~\cite{prl_kirill,bogdanov}. Such a mechanism of dramatic suppression of radiative losses for one of the interacting leaky modes is identical to emergence of accidental BICs in unbound structures, which originate in accord with the Friedrich-Wintgen scenario, when the radiating tails of several leaky modes cancel each other out via destructive interference resulting in a growth of the $Q$ factor.
\begin{figure}[t]
   \centering
\centering\includegraphics[width=0.95\linewidth]{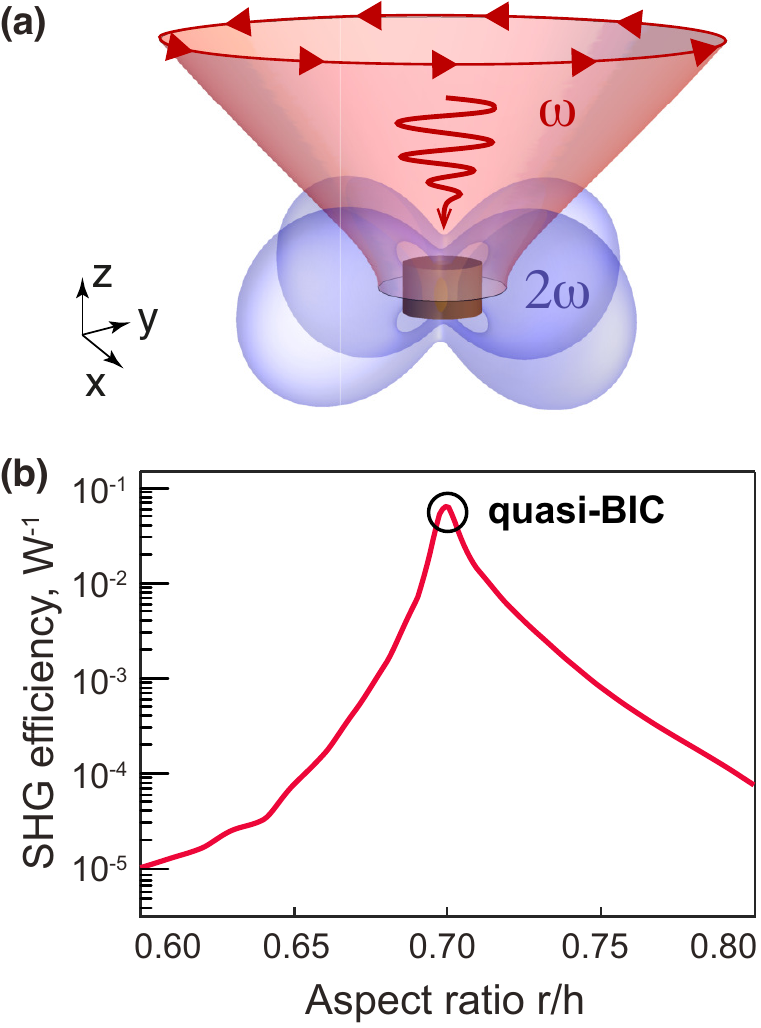}
 \caption{ {\bf Enhanced SHG in the quasi-BIC regime.} (a) Artistic view of the second-harmonic field generated from a subwavelength AlGaAs nanoantenna pumped by an azimuthally polarized beam at the BIC conditions.
(b) Dependence of the intrinsic SHG conversion efficiency on the nanoparticle aspect ratio $r/h$ with a large growth in the vicinity of the quasi-BIC~\cite{PRL_luca}. An open circle corresponds to the maximum of the SHG efficiency, and it coincides with the conditions of the quasi-BIC regime.}
\label{fig2}
\end{figure}

Figure~\ref{fig1} summarizes the physics of the quasi-BIC formation in isolated subwavelength high-index resonators. The geometry of a nanoparticle is shown in Fig.~\ref{fig1}(a), being a nanodisk made of Al$_{0.2}$Ga$_{0.8}$As suspended in a homogeneous background with refractive index 1. The values of permittivity and loss of AlGaAs are imported from the tabulated data~\cite{AlGaAs} for the wavelength range around $1600$~nm.

The eigenmodes of a finite-length nanodisk can be roughly divided into two families, namely, radially oscillating and axially oscillating modes. Hereinafter, we call radial resonances originating from resonances of an infinitely long cylinder as Mie-like modes and axial modes - as Fabry-P\' erot-like (FP-like) modes. A pair of low-frequency FP-like and Mie-like modes undergoes strong coupling via continuous tuning of the disk aspect ratio $r/h$ producing the characteristic avoided resonance crossing feature, as shown in Fig.~\ref{fig1}(b). Since the nanodisk represents an open resonator, the radiative lifetime of interacting modes is also strongly modified~\cite{WiersigPRL} leading to the generation of a quasi-BIC with the spectacular enhancement of $Q$ factor up to 120. This value is one order of magnitude higher than the $Q$ factor of conventional dipolar resonances in the same frequency range.

The electric field profile and far-field distributions of the high-$Q$ mode for the aspect ratio in- and out of the supercavity regime [see cases A, B, and C in Fig.~\ref{fig1}(b)] are shown in Fig.~\ref{fig1}(c). It is clearly seen from the near-field distributions that a Mie-like mode smoothly transforms to a FP-like mode through a quasi-BIC with a hybrid field distribution. The lower panel of Fig.~\ref{fig1}(c) shows that away from the quasi-BIC condition the far\textcolor{black}{-}field distribution of the high-$Q$ mode is dominated by the magnetic dipole contribution. Since both FP-like and Mie-like modes are characterized by the magnetic dipole radiation pattern, their hybridization leads to sufficient suppression of the magnetic dipole component enabling the next allowed multipole \textcolor{black}{--} a highly symmetric magnetic octupole term. Such a behaviour is in full agreement with the results of multipolar decomposition of the eigenmodes~\cite{bogdanov, bic_wei}, which explains the physical nature of the $Q$ factor enhancement.

The supercavity regime can be observed in the evolution of the scattering spectra with respect to the change of the aspect ratio of a high-index dielectric resonator which reveals the characteristic avoided crossing behaviour~\cite{prl_kirill}. The predicted features were recently confirmed in the proof-of-concept microwave experiment with a plastic cylindrical vessel filled with water~\cite{bogdanov}. The reported Rabi splitting was about $100$ MHz at the frequency of $1$ GHz.

In addition to the ARC hallmark, the quasi-BIC manifests itself as an abrupt change of the lineshape of the high-$Q$ mode. In particular, in a narrow vicinity of the supercavity mode, the conventional Fano profile transforms into a symmetric Lorentzian peak. It results in a maximum of the scattering cross-section which is exactly opposite to the so-called anapole regime when the scattering becomes suppressed~\cite{anapole}. This transition can be analytically described by the peculiarities of the Fano asymmetry parameter which has different signs before and after the quasi-BIC condition, and it diverges when the quasi-BIC condition is met~\cite{bogdanov}. Such a behavior is analogous to manifestation of BICs in unbound periodic structures, where the Fano parameter $q$ becomes ill-defined corresponding to the ``collapse'' of Fano resonance~\cite{Fonda1963}. Since the asymmetry of the Fano resonance is easily distinguishable without fitting, a quasi-BIC can be recognized in the experimental spectra. Thereby, the scattering spectra represent an unambiguous indicator of the supercavity regime in isolated subwavelength nanoparticles giving a deeper insight into the physics of BICs at the nanoscale.

A giant enhancement of the $Q$ factor of dielectric nanoresonators is of a paramount interest for nonlinear optics, opening up new horizons for active and passive nanoscale metadevices. The reason is that nonlinear optics at the nanoscale is governed by strong field confinement and resonant response~\cite{nonlinear_review}, and is not limited by phase matching~\cite{martti}. As a result, the processes of high-order harmonic generation can be boosted dramatically since they scale as $(Q/V)^n$, where $n$ is the process order and $V$ is the mode volume~\cite{mod_vol}. Therefore, the giant increase of the generation efficiency is expected in isolated subwavelength nanoantennas tuned to the BIC regime which was recently predicted for the case of the second-harmonic generation (SHG) in AlGaAs nanodisks~\cite{PRL_luca}.
\begin{figure*}[t]
   \centering
\centering\includegraphics[width=0.9\linewidth]{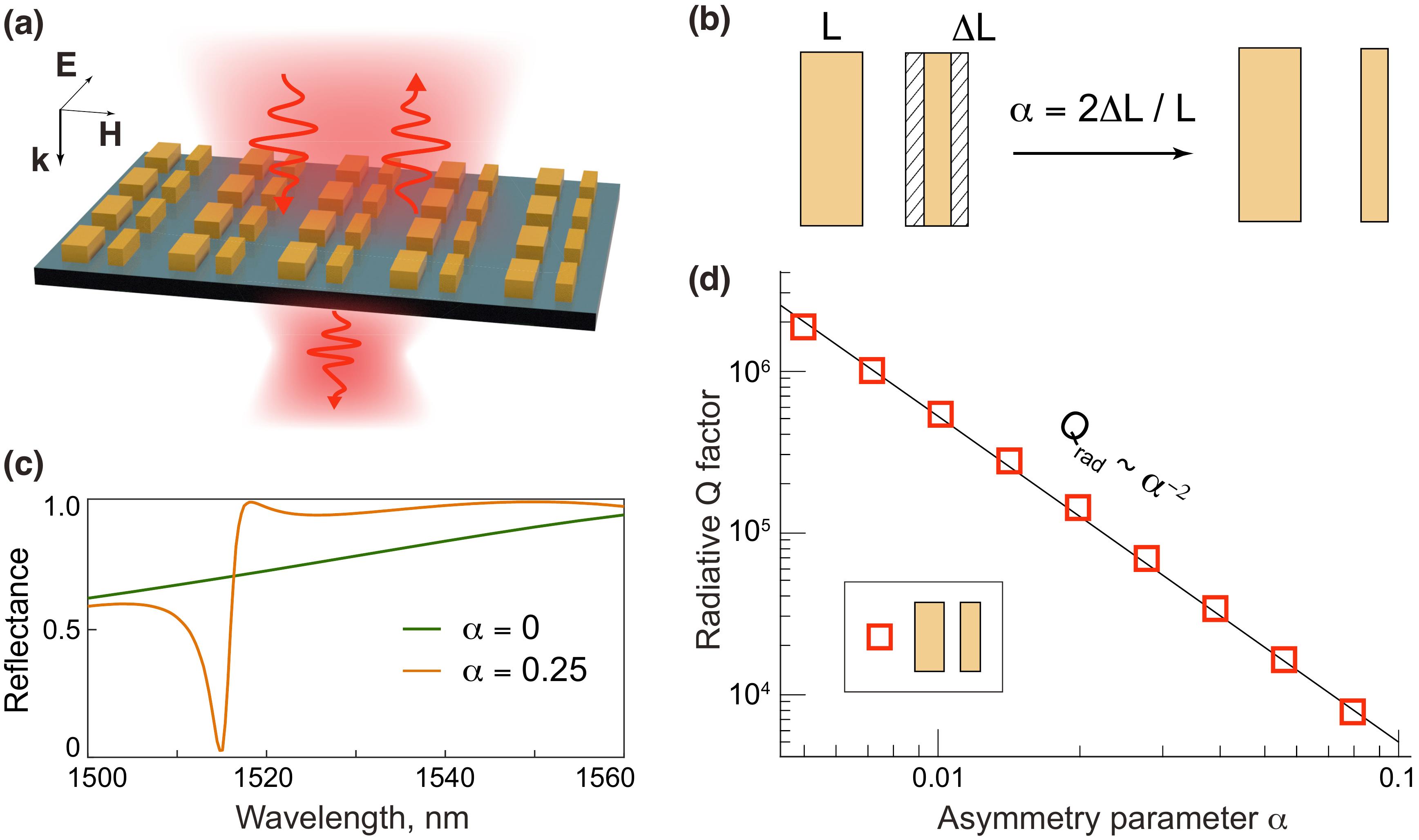}
 \caption{ {\bf Quasi-BIC in metasurfaces with a broken symmetry.} (a) Schematic of the scattering of light by an asymmetric metasurface composed of a square lattice of pairs of AlGaAs dielectric bars of different widths. Parameters are: period 950~nm, bar length 505~nm, bar width 250~nm, bar height 400~nm, and the distance between adjacent bars is 200~nm. (b) Definition of the asymmetry parameter $\alpha$. (c) Evolution of the reflectance spectrum of the metasurface from the symmetric geometry ($\alpha=0$) to a broken-symmetry geometry ($\alpha=0.25$). (d) Inverse quadratic dependence of the $Q$ factor on the asymmetry parameter $\alpha$ (log-log scale)~\cite{new_PRL}. The insert shows the specific shape of the isolated asymmetric meta-atom. }
\label{fig3}
\end{figure*}

Figure~\ref{fig2} outlines the results of our numerical analysis on the enhancement on the SHG efficiency for individual dielectric nanodisks in the supercavity regime. As shown in Fig~\ref{fig2}(a), the resonator is excited by an azimuthally polarized Gaussian beam providing perfect mode matching with the high-$Q$ resonance, whose electric field is uniform with respect to the azimuthal direction [see Fig.~\ref{fig1}(c)]. The geometrical and material parameters of the disk are the same as were for the analysis of the linear response. For calculations, we assume the value of the beam waist equal to $1.5~\mu$m and the value of the second-order nonlinear susceptibility equal to $290$ pm/V~\cite{susceptibility}.

Figure~\ref{fig2}(b) shows the intrinsic SHG conversion efficiency as a function of the disk aspect ratio, which is defined as $P_{\rm SH}/P_{0}^2$, where $P_{\rm SH}$ is the total power radiated at the second-harmonic frequency and $P_{0}$ is amount of the pump power actually coupled to the resonator. As it is shown, the SHG efficiency exhibits very dramatic growth in the vicinity of the supercavity mode, and it reaches its maximum exactly at the quasi-BIC condition. The predicted record-high value of the conversion efficiency for nanoscale resonators exceeds by two orders of magnitude the conversion efficiency observed at the magnetic dipole Mie resonance~\cite{PRL_luca}.

In addition, we also notice that high-$Q$ factors at the nanoscale can boost not only the sum-frequency generation efficiency, but also other processes, in particular, the spontaneous parametric downconversion. Recently, exploitation of the high-$Q$ supercavity mode of the AlGaAs nanodisk was shown to produce a strong enhancement of generation of entangled photon pairs associated with electric and magnetic dipole modes~\cite{Poddubny2018}.

\section{Metasurfaces}

Metasurfaces have attracted a lot of attention in the last years offering novel ways for wavefront shaping and advanced light focusing in combination with being ultra-thin optical elements~\cite{capasso,kruk}. Recently, metasurfaces based on high-index resonant dielectric materials have emerged as essential building blocks for various functional meta-optics devices~\cite{kruk2017functional} due to their low intrinsic loss, with unique capabilities for controlling the propagation and localization of light. A key concept underlying the specific functionalities of many metasurfaces is the use of constituent elements with spatially varying optical properties and optical response characterized by high $Q$ factors of the resonances.
\begin{figure}[t]
   \centering
\centering\includegraphics[width=0.9\linewidth]{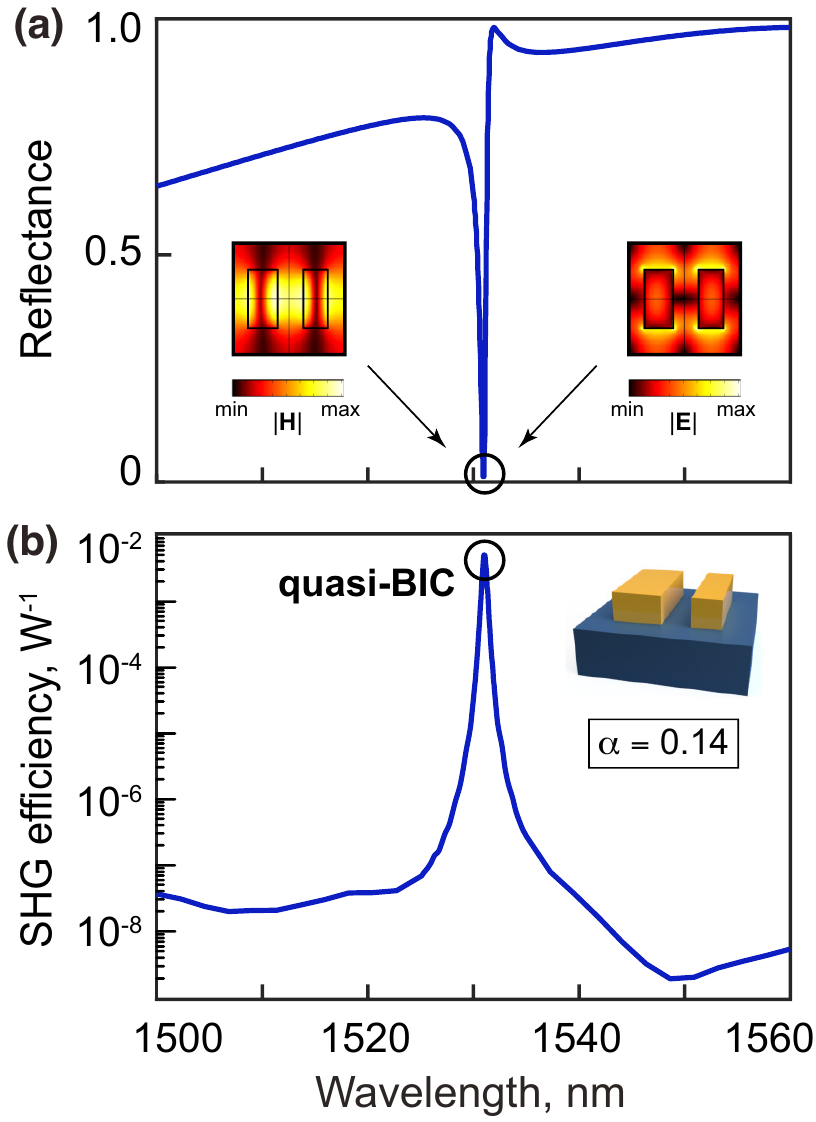}
 \caption{ {\bf Nonlinear metasurfaces.} Enhancement of the SHG conversion efficiency in broken-symmetry dielectric metasurfaces in the quasi-BIC regime. (a)~Linear reflectance spectrum of the broken-symmetry metasurface with the parameters as in Fig.~\ref{fig3} and the difference in bar widths equal to 35~nm. The insets show the distribution of electric and magnetic fields of the excited quasi-BIC within the unit cell. (b) Dependence of the intrinsic SHG conversion efficiency on the pump wavelength calculated by nonlinear simulations. The open circle shows that the maximum of the SHG efficiency coincides with the quasi-BIC regime. The inset shows the specific shape of the metasurface unit cell.}
\label{fig4}
\end{figure}

Recently, the resonant response of a special class of all-dielectric metasurfaces composed of meta-atoms with broken in-plane inversion symmetry was shown to demonstrate sharp resonances in the normal incidence reflection and transmission measurements.  Such sharp Fano features are associated with modes possessing large $Q$ factors which result in various unusual phenomena, including efficient imaging-based molecular barcoding~\cite{science} and enhanced nonlinear response~\cite{brener}. In a very recent paper it was proven that all high-$Q$ resonances in such seemingly different metasurfaces with asymmetric unit cells can be unified by the BIC concept~\cite{new_PRL}. In particular, it was validated that each sharp resonance originates from a symmetry-protected BIC which transforms to a quasi-BIC due to breaking of the in-plane inversion symmetry. More specifically, it was shown that the unit cell asymmetry induces an imbalance of the interference between contra-propagating leaky waves comprising the symmetry-protected BIC that leads to radiation leakage.

Figure~\ref{fig3} combines the main results on the quasi-BIC formation in metasurfaces with broken in-plane inversion symmetry. Aiming further analysis of nonlinear resonant properties, we propose an original design of a square lattice metasurface with parallel dielectric bars of a slightly different width, as shown in Fig.~\ref{fig3}(a). The metasurface is made of Al$_{0.3}$Ga$_{0.7}$As, and it is placed on a fused quartz substrate; all material properties include losses being imported from the tabulated data~\cite{AlGaAs}. Figure~\ref{fig3}(b) introduces the definition of the asymmetry parameter $\alpha$. Nonzero difference in bar widths results in emergence of a sharp Fano profile in the reflectance spectrum [see results of numerical calculations in Fig.~\ref{fig3}(c)]. Excitation beam is a normally incident plane wave linearly polarized along the bars. Figure~\ref{fig3}(d) shows that the linewidth of such a resonance and, consequently, the mode $Q$ factor is straightforwardly controlled by the magnitude of $\alpha$ defined in Fig.~\ref{fig3}(b).

Remarkably, for relatively small values of $\alpha$ the dependence of the $Q$ factor on the asymmetry parameter is inverse quadratic which is the universal law for all metasurfaces with broken in-plane symmetry of meta-atoms~\cite{new_PRL}. For structures with weak dispersion of material properties the operating wavelength and the $Q$ factor of a resonant symmetry-broken metasurface can be controlled efficiently by tuning the asymmetry parameter combined with a geometric rescaling of the structure. As a result, the understanding of the BIC-inspired nature of high-$Q$ resonances paves the way towards smart engineering of metasurfaces-on-demand required for various applications of linear and flat optics.

Similarly to the case of individual high-$Q$ nanoantennas, the enlargement of $Q$ factor for metasurfaces boosts the conversion efficiency of sum-frequency generation processes dramatically. Below we present original results on the SHG efficiency for an example of broken-symmetry all-dielectric metasurface. We focus on the design shown in Fig.~\ref{fig3} with the difference between bar widths equal to 35~nm ($\alpha=0.14$) which results in quasi-BIC with the $Q$ factor of 2500. In practice, it is not necessary to make $Q$ factor more than $f/\delta f$, where $f$ and $\delta f$ are the center frequency and the bandwidth of the incident pulse, respectively. In calculations we assume the value of $290$ pm/V for the second-order nonlinear susceptibility~\cite{susceptibility}.

The results for the enhancement of the SHG conversion efficiency in the quasi-BIC regime are summarized in Fig.~\ref{fig4}. We again focus on the intrinsic conversion efficiency defined as for the case of isolated nanoparticles. The reflectance spectrum manifests a sharp Fano resonance at the wavelength of about 1530~nm, as shown in Fig.~\ref{fig4}(a). The insets show the in-plane cross-sections of the near-field distribution of the quasi-BIC. Figure~\ref{fig4}(b) demonstrates a rapid growth of the SHG conversion efficiency as a function of the pump wavelength with a pronounced maximum at the wavelength of the quasi-BIC. Further enhancement of the conversion efficiency is possible for even smaller values of the asymmetry parameter being limited only by the material losses and structural imperfections.

\section{Conclusion and outlook}

Resonant dielectric nanoparticles and metasurfaces provide an alternative solution to enhance the performance of many nanophotonic structures and metadevices employing metallic elements and plasmonics effects. The study of resonant dielectric meta-optics has become a new research direction in metamaterials and nanophotonics, and it is expected to complement or substitute some of the plasmonic components in a range of potential applications.  The unique low-loss resonant behavior makes it possible to reproduce many subwavelength resonant effects already demonstrated in nanoscale plasmonics without essential energy dissipation.

However, dielectric nanostructures supporting optically-induced Mie-type electric and magnetic resonances open many novel opportunities associated with interference effects and strong mode coupling, in both isolated high-index dielectric photonic resonators and all-dielectric metasurfaces.  The co-existence of strong electric and magnetic resonances, their interference, and the resonant enhancement of the magnetic response in dielectric nanoparticles bring new physics and entirely novel functionalities to simple geometries; a direction that has not been explored much so far, especially in the nonlinear regime.

A novel approach is associated with the bound states in the continuum which originate from strong coupling between the modes in resonant dielectric  structures. They attracted a lot of attention in photonics recently as an alternative mean to achieve very large $Q$ factors for lasing and also tune the system to the quasi-BIC supercavity regime. Importantly, even a single subwavelength high-index dielectric resonator can be tuned into the regime of a supercavity mode. This can be achieved by varying the nanoparticle's aspect ratio, when the radiative losses are almost suppressed due to the Friedrich-Wintgen scenario of destructive interference of leaky modes. In connection with broader applications, many metasurfaces composed of arrays of meta-atoms with broken in-plane symmetry can support high-$Q$ resonances, which are directly associated with the concept of bound states in the continuum. This includes, in particular, broken-symmetry Fano dielectric metasurfaces for enhancement of nonlinear effects and recently reported pixelated dielectric metasurfaces for sensing. All such structures can be employed in nonlinear nanophotonics, since they give rise to strong light-matter interaction on the nanoscale.

We believe that the novel approach to employ the BIC physics for enhancing the resonances in all-dielectric subwavelength structures and metasurfaces, and a recently proven direct link between the Fano resonances and quasi-BICs, provides a powerful tool to engineer the high-$Q$ resonances and Purcell factors in nanoscale metadevices, allowing their numerous applications in sensing, nonlinear optics, quantum physics, and active nanophotonics.

\begin{acknowledgments}
The authors thank Prof. Cheng-Wei Qiu for his kind invitation to contribute to this special issue, and they acknowledge productive collaboration and useful discussions with many colleagues and students, especially A. Al\'u, S. Kruk, S. Lepeshov, M. Limonov, Mingkai Liu, Wei Liu, Hong-Gyu Park, M. Rybin, D. Smirnova, and I. Staude. The authors acknowledge a financial support from the Australian Research Council, the Strategic Fund of the Australian National University, the Alexander von Humboldt Foundation, and the Russian Science Foundation (grant 18-72-10140). K. K. and A. B. acknowledge the support from the Foundation for the Advancement of Theoretical Physics and Mathematics "BASIS" (Russia).
\end{acknowledgments}


\begin{thebibliography}{99}

\bibitem{kuznetsov2016optically}
A. I. Kuznetsov, A. E. Miroshnichenko, M. L. Brongersma, Y. S. Kivshar, and B. Luk'ayanchuk,
{\it Optically resonant dielectric nanostructures},
Science {\bf 354}, aag2472 (2016).


\bibitem{staude}
I. Staude and J. Schilling,
{\it Metamaterial-inspired silicon nanophotonics},
Nature Photonics {\bf 11}, 274 (2017).

\bibitem{kruk2017functional}
S. Kruk and Y. Kivshar,
{\it Functional meta-optics and nanophotonics governed by Mie resonances},
ACS Photonics {\bf 4}, 2638 (2017).

\bibitem{nmat}
N. Zheludev, and Yu.,S. Kivshar,
{\it From metamaterials to metadevices},
Nature Materials {\bf 11}, 917â924 (2012).

\bibitem{Moitra2015}
P. Moitra, B. A. Slovick, W. Li, I. I. Kravchencko, D. P. Briggs, S. Krishnamurthy, and J. Valentine,
{\it Large-scale all-dielectric metamaterial perfect reflectors},
ACS Photon. {\bf 2}, 692 (2015).

\bibitem{Faraon2015}
A. Arbabi, Y. Horie, M. Bagheri, and A. Faraon,
{\it Dielectric metasurfaces for complete control of phase and polarization with subwavelength spatial resolution and high transmission},
Nat. Nanotechnol. {\bf 10}, 937 (2015).

\bibitem{Shalaev2015}
M. I. Shalaev, J. Sun, A. Tsukernik, A. Pandey, K. Nikolskiy, and N. M. Litchinitser,
{\it High-efficiency all-dielectric metasurfaces for ultracompact beam manipulation in transmission mode},
Nano Lett. {\bf 15}, 6261 (2015).

\bibitem{Yu2015}
Y. F. Yu, A. Y. Zhu, R. Paniagua-Dominguez, Y. H. Fu, B. Lukyanchuk, and A. I. Kuznetsov,
{\it High-transmission dielectric metasurface with 2$\pi$ phase control at visible wavelengths},
 Laser Photon. Rev. {\bf 9}, 412 (2015).

\bibitem{yuri}
Y.S. Kivshar, {\it All-dielectric meta-optics and nonlinear nanophotonics},
National Science Review {\bf 5}, 144 (2018).

\bibitem{oe_wei}
W. Liu and Y. Kivshar,
{\it Generalized Kerker effects in nanophotonics and meta-optics},
Optics Express {\bf 26}, 13085 (2018).

\bibitem{von1929uber}
J. Von Neuman and E. Wigner,
{\it \"{U}ber Merkw\"{u}rdige Diskrete Eigenwerte},
Phys. Z. {\bf 30}, 467 (1929).


\bibitem{Fonda1963}
L. Fonda,
{\it Bound states embedded in the continuum and the formal theory of scattering},
Ann. Phys. {\bf 22}, 123 (1963).

\bibitem{Stillinger1960}
F. H. Stillinger and D. R. Herrick,
{\it Bound states embedded in the continuum and the formal theory of scattering},
Phys. Rev. A {\bf 11}, 446 (1975).

\bibitem{fri_win}
H. Friedrich and D. Wintgen,
{\it Interfering resonances and bound states in the continuum},
Phys. Rev. A {\bf 32}, 3231 (1985).


\bibitem{Ursell1951}
F. Ursell,
{\it Trapping modes in the theory of surface waves},
Math. Proc. Cambridge Philos. Soc. {\bf 47}, 347 (1951).


\bibitem{Cumpsty1971}
N.\,A. Cumpsty and  D.\,S. Whitehead,
{\it The excitation of acoustic resonances by vortex shedding},
J. Sound Vib. {\bf 18}, 353 (1971).

\bibitem{Parker1989}
R. Parker and  S.\,A. Stoneman,
{\it The excitation and consequences of acoustic resonances in enclosed fluid flow around solid bodies},
Proc. Inst. Mech. Eng. C  {\bf 203}, 9 (1989).



\bibitem{BIC_review}
C.W. Hsu, B. Zhen, A.D. Stone, J.D. Joannopoulos, and M. Soljaci\'c,
{\it Bound states in the continuum},
Nature Review Materials {\bf 1}, 16048 (2016).


\bibitem{Paddon2000}
P. Paddon and J.\,F. Young
{\it Two-dimensional vector-coupled-mode theory for textured planar waveguides},
Phys. Rev. B {\bf 61}, 2090 (2000).

\bibitem{Tikhodeev2002}
S.\,G. Tikhodeev, A.\,L. Yablonskii, E.\,A. Muljarov, N.\,A. Gippius, and T. Ishihara,
{\it Quasiguided modes and optical properties of photonic crystal slabs},
Phys. Rev. B {\bf 66}, 045102 (2002).

\bibitem{Shipman2005}
S.\,P. Shipman and S. Venakides,
{\it Resonant transmission near nonrobust periodic slab modes},
Phys. Rev. E {\bf 71}, 026611 (2005).

\bibitem{Bulgakov2014}
E.\,N. Bulgakov and A.\,F. Sadreev,
{\it Robust bound state in the continuum in a nonlinear microcavity embedded in a photonic crystal waveguide},
Opt. Lett. {\bf 17}, 5212 (2014).

\bibitem{Bulgakov2017}
E. Bulgakov, A. F. Sadreev, and D. Maksimov,
{\it Light trapping above the light cone in one-dimensional srrays of dielectric spheres},
Appl. Sci {\bf 7}, 147 (2017).

\bibitem{Belyakov2018}
M.\,A. Belyakov, M.\,A. Balezin, Z.\,F. Sadrieva, P.\,V. Kapitanova, E.\,A. Nenasheva, A.\,F. Sadreev, and A.\,A. Bogdanov,
{\it Experimental observation of symmetry protected bound state in the continuum in a chain of dielectric disks},
arXiv:1806.01932 (2018).


\bibitem{Marinica2008}
D.\,C. Marinica, A.\,G. Borisov,  and S.\,V. Shabanov,
{\it Bound states in the continuum in photonics},
 Phys. Rev. Lett. {\bf 100}, 183902 (2008).

\bibitem{Ndangali2010}
R.\,F. Ndangali and S.\,V. Shabanov,
{\it Electromagnetic bound states in the radiation continuum for periodic double arrays of subwavelength dielectric cylinders},
J. Math. Phys. {\bf 51}, 102901 (2010).

\bibitem{Bulgakov2015}
E.\,N. Bulgakov, K.\,N. Pichugin, and A.\,F. Sadreev,
{\it All-optical light storage in bound states in the continuum and release by demand},
Opt. Express {\bf 23}, 22520 (2015).


\bibitem{GomisBresco2017}
J. Gomis-Bresco, D. Artigas, and L. Torner,
{\it Anisotropy-induced photonic bound states in the continuum},
Nat. Photonics {\bf 11}, 232 (2017).

\bibitem{Zhong2017}
H.-H. Zhong, Z. Zhou, B. Zhu, Y.-G. Ke, and C.-H. Lee,
{\it Floquet Bound states in a driven two-particle Bose-Hubbard model with an impurity},
Chinese Phys. Lett. {\bf 34}, 070304 (2017).

\bibitem{Longhi2014}
S. Longhi,
{\it Bound states in the continuum in PT-symmetric optical lattices},
Opt. Lett. {\bf 39}, 1697 (2014).

\bibitem{Kartashov2018}
Y.\,V Kartashov, C. Milian, V.\,V. Konotop, and L. Torner,
{\it Bound states in the continuum in a two-dimensional PT-symmetric system},
Opt. Lett. {\bf 43}, 575 (2018).

\bibitem{Foley2014}
J.\,M. Foley, S.\,M. Young, and J.\,D. Phillips,
{\it Symmetry-protected mode coupling near normal incidence for narrow-band transmission filtering in a dielectric grating},
Phys. Rev. B {\bf 89}, 165111 (2014).

\bibitem{Liu2017}
Y. Liu, Z. Weidong, and S. Yuze,
{\it Optical refractive index sensing based on high-$Q$ bound states in the continuum in free-space coupled photonic crystal slabs},
Sensors {\bf 17}, 1861 (2017).

\bibitem{Romano2018}
S. Romano, A. Lamberti, M. Masullo, E. Penzo, S. Cabrini, I. Rendina, and V. Mocella,
{\it Optical biosensors based on photonic crystals supporting bound states in the continuum},
Materials {\bf 11}, 526 (2018).

\bibitem{BIC_nature}
A. Kodigala, Th. Lepetit, Q. Gu, B. Bahari, Y. Fainman, and B. Kant\'e,
{\it Lasing action from photonic bound states in continuum},
Nature {\bf 541}, 196â199 (2017)

\bibitem{kuznetsov_lasing}
S. T. Ha, Y. H. Fu, N. K. Emani, Z. Pan, R. M. Bakker, R. Paniagua-Domanguez, and A. I. Kuznetsov,
{\it Directional lasing in resonant semiconductor nanoantenna arrays},
Nat. Nanotechnology {\bf 1}, (2018).

\bibitem{sensing}
S. Romano, G. Zito, S. Torino, G. Calafiore, E. Penzo, G. Coppola, S. Cabrini, I. Rendina, and V. Mocella,
{\it Label-free sensing of ultralow-weight molecules with all-dielectric metasurfaces supporting bound states in the continuum},
Photonics Research {\bf 6}, 726 (2018).

\bibitem{SERS}
S. Romano, G. Zito, S. Manag\'o, G. Calafiore, E. Penzo, S. Cabrini, A. C. De Luca, and V. Mocella,
{\it Surface-Enhanced Raman and Fluorescence Spectroscopy with an All-Dielectric Metasurface},
 J. Phys. Chem. C {\bf 122}, 19738 (2018).




\bibitem{Krasikov2018}
H.-H. Zhong, Z. Zhou, B. Zhu, Y.-G. Ke, C.-H. Lee,
{\it Nonlinear bound states in the continuum of a one-dimensional photonic crystal slab},
Phys. Rev. B {\bf 97}, 224309 (2018).


\bibitem{Bulgakov2017_1}
E.\,N. Bulgakov and A.\,F. Sadreev,
{\it Bound states in the continuum with high orbital angular momentum in a dielectric rod with periodically modulated permittivity},
Phys. Rev. A {\bf 96}, 013841 (2017).

\bibitem{Bulgakov2017_2}
E.\,N. Bulgakov and A.\,F. Sadreev,
{\it Trapping of light with angular orbital momentum above the light cone},
Adv. Electromagn. {\bf 6}, 1 (2017).

\bibitem{BIC_exciton}
K. L. Koshelev,  S. K. Sychev, Z. F. Sadrieva, A. A. Bogdanov, and I. V. Iorsh,
{\it Strong coupling between excitons in transition metal dichalcogenides and optical bound states in the continuum},
Phys. Rev. B  (2018).

\bibitem{Poddubny2018}
A.\,N. Poddubny, D.\,A. Smirnova,
{\it Nonlinear generation of quantum-entangled photons from high-$Q$ states in dielectric nanoparticles},
arXiv:1808.04811 (2018).






\bibitem{Sadrieva2017}
Z.F. Sadrieva, I.S. Sinev, K.L. Koshelev, A. Samusev, I.V. Iorsh, O. Takayama, R. Malureanu, A.A. Bogdanov, and A.V. Lavrinenko,
{\it Transition from optical bound states in the continuum to leaky resonances: Role of substrate and roughness},
ACS Photonics {\bf 4}, 723 (2017).


\bibitem{prl_kirill}
M.V. Rybin, K.L. Koshelev, Z.F. Sadrieva, K.B. Samusev, A.A. Bogdanov, M.F. Limonov, and Yu.S. Kivshar,
{\it High-$Q$ supercavity modes in subwavelength dielectric resonators},
Phys. Rev. Lett. {\bf 119}, 243901 (2017).

\bibitem{rybin}
M. Rybin and Yu.S. Kivshar,
{\it Supercavity lasing},
Nature {\bf 541}, 165 (2017).

\bibitem{Taghizadeh2017}
A. Taghizadeh, and I.-S. Chung,
{\it Quasi bound states in the continuum with few unit cells of photonic crystal slab},
Appl. Phys. Lett. {\bf 111}, 031114 (2017).

\bibitem{new_PRL}
K.~Koshelev, S.~Lepeshov, M.~Liu, A.~Bogdanov, and Y.S.~Kivshar,
{\it Asymmetric metasurfaces and high-$Q$ resonances governed by bound states in the continuum},
submitted to Phys. Rev. Lett. (2018); arXiv:1809.00330.

\bibitem{brener}
S. Campione, S. Liu, L.I. Basilio, L.K. Warne, W.L. Langston, T.S. Luk, J.R. Wendt, J.L. Reno, G.A. Keeler, I. Brener, and M.B. Sinclair,
{\it Broken symmetry dielectric resonators for high quality factor Fano metasurfaces},
ACS Photonics {\bf 3}, 2362 (2016).

\bibitem{science}
A. Tittl, A. Leitis, M. Liu, F. Yesilkoy, D.Y. Choi, D.N. Neshev, Y.S. Kivshar, and H. Altug,
{\it Imaging-based molecular barcoding with pixelated dielectric metasurfaces},
Science {\bf 360}, 1105 (2018).

\bibitem{active}
S. Han, L. Cong, Y. K. Srivastava, B. Qiang, M. V. Rybin, W. X. Lim, Q. Wang, Yu. S. Kivshar, and R. Singh,
{\it All-dielectric active photonics driven by bound states in the continuum},
arXiv preprint arXiv:1803.01992 (2018).


\bibitem{bogdanov}
A.A. Bogdanov, K.L. Koshelev, P.V. V. Kapitanova, M.V. Rybin, S.A. Gladyshev, Z.F. Sadrieva, K.B. Samusev, Yu.S. Kivshar, and M.F. Limonov,
{\it A direct link between Fano resonances and bound states in the continuum},
arXiv preprint, arXiv:1805.09265 (2018).

\bibitem{AlGaAs}
\url{https://refractiveindex.info/}

\bibitem{WiersigPRL}
J. Wiersig,
{\it Formation of long-lived, scarlike modes near avoided resonance crossings in optical microcavities},
Phys. Rev. Lett. {\bf 97}, 253901 (2006).

\bibitem{bic_wei}
Weijin Chen, Yuntian Chen, and Wei Liu,
{\it Subwavelength high-$Q$ Kerker supermodes with unidirectional radiations},
arXiv preprint, arXiv: 1808.05539v1 (2018).

\bibitem{anapole}
A. E. Miroshnichenko, A. B. Evlyukhin, Y. F. Yu, R. M. Bakker, A. Chipouline, A. I. Kuznetsov, B. Lukâyanchuk, B. N. Chichkov, and Yu. S. Kivshar,
{\it Nonradiating anapole modes in dielectric nanoparticles},
Nature Commun. {\bf 6}, 8069 (2015).

\bibitem{nonlinear_review}
D. Smirnova and Y.S. Kivshar,
{\it Multipolar nonlinear nanophotonics},
Optica {\bf 3}, 1241 (2016).

\bibitem{martti}
M. Kauranen,
{\it Freeing Nonlinear Optics from Phase Matching},
Science {\bf 342}, 1182 (2013).

\bibitem{mod_vol}
M. Notomi,
{\it Strong Light Confinement With Periodicity},
Proc. IEEE {\bf 99}, 1768 (2011).

\bibitem{PRL_luca}
L. Carletti , K. Koshelev , C. De Angelis , and Yu. Kivshar,
{\it Giant nonlinear response at the nanoscale driven by bound states in the continuum},
Phys. Rev. Lett. {\bf 121}, 033903 (2018).

\bibitem{susceptibility}
I. Shoji, T. Kondo, A. Kitamoto, M. Shirane, and R. Ito,
{\it Absolute scale of second-order nonlinear-optical coefficients},
J. Opt. Soc. Amer. B {\bf 14}, 2268 (1997).


\bibitem{capasso}
N. Yu and F. Capasso,
{\it Flat optics with designer metasurfaces},
Nat. Mater. {\bf 13}, 139 (2014).

\bibitem{kruk}
S. Kruk, B. Hopkins, I.I. Kravchenko, A. Miroshnichenko, D.N. Neshev, and Y.S. Kivshar
{\it Broadband highly efficient dielectric metadevices for polarization control},
APL Photonics {\bf 1}, 030801 (2016).


\end{thebibliography}
\end{document}